\documentclass[prl,preprint,superscriptaddress]{revtex4}

\include{psfig}

\begin{document}

\title{\bf A new expansion around mean field for the quantum
Ising model}

\author{F. de Pasquale}
\address{Departement of Physics University of Rome "La Sapienza", P-le
A.Moro 2, 00185 Rome Italy}
\address{\mbox{$^b$ INFM Rome}}
\author{S.M. Giampaolo}
\address{Departement of Physics University of Rome "La Sapienza", P-le
A.Moro 2, 00185 Rome Italy}
\date{Nov. 23 2001}

\begin{abstract}
We show that an high temperature expansion at fixed order
parameter can be derived for the quantum Ising model. The basic
point is to consider a statistical generating functional
associated to the local spin state. The probability at thermal
equilibrium of this state reflects directly the occurrence of a
spontaneous symmetry breaking. It is possible to recover the
expansion around the mean field in the system dimensionality if
the ``direction'' in the Hilbert space of local spin states is
suitably chosen.  Results for the free energy at the critical
temperature, as a function of the transverse field, in first order
approximation in the inverse system dimensionality are compared
with those of the standard approach.
\end{abstract}
\pacs{}

\maketitle

The recent interest in decoherence\cite{zurek} phenomena in
connection with quantum computing leads naturally to the study of
a local spin state interacting with an environment. It has been
shown that quantum computation can be performed acting on a
nuclear spin at thermal equilibrium\cite{jones}. In this
perspective it is interesting to study the simplest quantum spin
model, i.e the Ising model in a transverse field. This model has
an exact solution in one dimension\cite{mattis}, and approximate
techniques, such as the random phase approximation \cite{wang} and
exact cumulant series expansion\cite{pfeuty,weihong} have also be
applied for higher order. The main result is due to Suzuki's
work\cite{suzuki} who showed the correspondence with a classical
model in d+1 dimensions. A systematic expansion around the mean
field result is however still lacking. As we shall see such an
expansion can be introduced if the phase transition is
characterized by a statistical generating functional associated to
the local spin state. The probability of finding the spin in the
i-th site in the general state $|s_i>=\cos
\frac{\theta_i}{2}|\uparrow_i>+e^{i\varphi_i}\sin
\frac{\theta_i}{2}|\downarrow_i>$ when the system is in thermal
equilibrium is
\begin{equation}
\label{def1}
 \rho_i(\theta_i,\varphi_i)=\frac{Tr\left[e^{-\beta
H}|s_i><s_i|\right]} {Tr \left(e^{-\beta H}\right)}
\end{equation}
It is convenient to express the projection operators in terms of
the Pauli spin operators
\begin{equation}
\label{def2}
 |s_i><s_i|=\frac{1}{2}+S_i \cdot n_i
\end{equation}
Here $n_i$
$(\sin\theta_i\cos\varphi_i,\sin\theta_i\sin\varphi_i,\cos\theta_i)$
and $S_i$ $(S_i^x,S_i^y,S_i^z)$ are respectively the unitary
vector which defines the symmetry breaking direction, and the spin
- $\frac{1}{2}$ vector operator. Taking into account Eq.
(\ref{def2}), the local spin state probability becomes
\begin{equation}
\label{def3}
 \rho_i(\theta_i,\varphi_i)=\frac{1}{2} + \frac{Tr
  \left[ \exp(-\beta H) S_i \cdot n_i\right]}{Tr \left[ \exp(-\beta H) \right] }
\end{equation}
Note that the non trivial local spin probability $\rho_i -
\frac{1}{2}$ corresponds to the magnetization in the symmetry
breaking direction. It is immediately seen that $\rho_i$ is
directly related to the symmetry breakdown. In the quantum Ising
model
\begin{equation}
 \label{Hamiltonian}
H=-\sum_{i,j}J_{i,j}S_{i}^{z}S_{j}^{z}-\sum_ih_iS_{i}^{x}
\end{equation}
the only symmetry is the spin inversion $S_i^z \rightarrow
-S_i^z$. Above the critical temperature $T_c$, the magnetization
is directed along the transverse field direction, while in the
ordered phase a non vanishing component in the z direction
appears. The local state probability of the symmetrical phase
(above $T_c$) is
\begin{equation}
\label{abovetc}
 \rho_i^S(\theta_i,\varphi_i)=\frac{1}{2} + \frac{Tr \left[
  \exp(-\beta H) S_i \cdot n_{i \bot} \right]}
  {Tr \left[ \exp(-\beta H) \right] }
\end{equation}
where $ n_{\bot i}$ is the unitary vector orthogonal to the z
direction. When the symmetry is broken (below $T_c$)
\begin{equation}
\label{undertc}
 \rho_i^{SB}(\theta_i,\varphi_i)=\rho_i^S(\theta_i,\varphi_i)+
 \frac{Tr \left[ \exp(-\beta H) S_i^z \cos \theta_i \right]}
 {Tr \left[ \exp(-\beta H) \right] }
\end{equation}
We see that the qualitative change of the local probability is
directly related, as expected, to a non vanishing statistical
average of $S_i^z$.
The non trivial part of the local spin state
probability  is the first moment of the statistical generating
function $G$ in the limit of vanishing source field $\lambda_i$.
\begin{equation}
\label{defG}
 G=\ln \left\{ Tr \left[e^{-\beta H}exp\left( \sum_i \lambda_i S_i
 \cdot n_i \right) \right] \right\}
\end{equation}
We see that for vanishing $\lambda_i$, $G$ becomes the free
energy. Moreover in the absence of a transverse field and for
$\theta_i=0$ the statistical functional $G$ reduced to that of the
Ising model in the presence of a symmetry breaking field along the
z - direction $h_i=\frac{\lambda_i}{\beta}$.

Expansion around the mean field are usually derived by introducing
a generalization of the free energy in the presence of symmetry
breaking fields and then by Legendre transform which makes the
order parameter the independent variable of the problem. This
generalization is usually accomplished introducing symmetry
breaking fields as perturbation in the system Hamiltonian. We
consider here an alternative generalization considering the
generating function of Eq. (\ref{defG}). The main advantage of the
present approach is the possibility of a straightforward expansion
in the inverse temperature. At infinite temperature there is a
complete degeneracy of the local spin state, which is removed only
for the presence of the symmetry breaking fields. At finite
temperature, the correction due to spin - spin interaction and the
transverse field compete each other to determine the local spin
state probability and the local magnetization.

Let us to introduce the Legendre transform of the generating
function $W(m)=G(\lambda)-\sum_i\lambda_i m_{i}$. Here
$m_i=<S_i\cdot n_i>$ and angular brackets stand for
\begin{equation}
 \label{defthermal}
 <\widehat{O}> =\frac{Tr\left[e^{-\beta H} \widehat{O} exp\left( \sum_i \lambda_i(\beta)
 ( S_i \cdot n_i)\right)\right]}
 {Tr\left[e^{-\beta H} exp\left( \sum_i  \lambda_i(\beta)
 ( S_i \cdot n_i) \right)\right]}
\end{equation}
Note that $m_i$, as a consequence of the relation between $m_i$
and the generating function $m_i=\frac{\partial G}{\partial
\lambda_i}$, become the independent variables of $W$. On the other
hand, the source field $\lambda_i$ depends on $m_i$ and $\beta$
\begin{equation}
 \label{lambdafuncm}
 \lambda_i= - \frac{\partial W}{\partial m_i}
\end{equation}
The free energy in the limit of vanishing source field is given by
$W$ calculated for the ``magnetization'' which satisfies extremum
condition $\frac{\partial W}{\partial m_i}=0$. This is the same
procedure which applies to the free energy of the Ising
model\cite{georges}.Following reference (2), we exploit the limit
of vanishing $\beta$ to fix the relation between $m_i$ and
$\lambda_i(0)$
\begin{equation}
 \label{invrelation}
  m_i= \frac{1}{2} \tanh \left( \frac{\lambda_i(0)}{2}\right)
\end{equation}
The next step is to derive the high temperature expansion of W.
The expansion up to the second order in $\beta$ is related to the
following quantities.
 \begin{widetext}
\begin{eqnarray}
 \label{coeff}
  W|_{\beta=0} & = & \sum_i \ln \left[ 2 \cosh
  \left(\frac{\lambda_i(0)}{2}\right)\right]-\lambda_i(0)m_i \nonumber \\
  \left.\frac{\partial W}{\partial \beta}\right|_{\beta=0} & = & <H>_{0} \\
  \left.\frac{\partial^2 W}{\partial \beta^2}\right|_{\beta=0} & = &
  <H^2>_{0}-(<H>_0)^2-2\sum_i\left. \frac{\partial \lambda_i}{\partial
  \beta}\right|_{\beta=0}<H(S_i\cdot n_i -m_i)>_0 + \nonumber \\
  & & \sum_{i,j}\left. \frac{\partial \lambda_i}{\partial
  \beta}\right|_{\beta=0}\left.\frac{\partial \lambda_j}{\partial
  \beta}\right|_{\beta=0}<(S_i\cdot n_i -m_i)(S_j \cdot n_j -m_j)>_0
  \nonumber
\end{eqnarray}
 \end{widetext}
 There the symbol $<>_0$ stands for the statistical
average defined in Eq. (\ref{defthermal}) evaluated at $\beta=0$,
and the derivative of $\lambda_i$ with respect to $\beta$ are
obtained from Eq. (\ref{lambdafuncm})
\begin{equation}
 \label{lambda}
  \left. \frac{\partial \lambda_i}{\partial \beta}
  \right|_{\beta=0}=-
  \left. \frac{\partial^2 W}{\partial \beta \partial m_i} \right|_{\beta=0}
\end{equation}
A straightforward calculation gives
 \begin{widetext}
\begin{eqnarray}
 \label{GeneralW}
 W & \approx & -\sum_i \left( \frac{1}{2}+m_i \right) \ln\left( \frac{1}{2}+m_i \right) +
    \left( \frac{1}{2}-m_i \right) \ln\left( \frac{1}{2}-m_i
    \right) \nonumber \\
     & & +\beta \sum_{i,j} J_{i,j}m_im_j \cos \theta_i \cos \theta_j +
   \beta \sum_{i} h_{i}m_i \sin \theta_i \cos \varphi_i  \nonumber \\
 & & +\frac{\beta^2}{8} \sum_{i,j,k}(J_{i,j}+J_{j,i})(J_{i,k}+J_{k,i})m_jm_k \cos\theta_k
  \cos \theta_j \sin^2 \theta_i + \frac{\beta^2}{8}\sum_ih_i^2(1-\sin^2 \theta_i \cos^2
  \varphi_i) \nonumber \\
  & & - \frac{\beta^2}{4}\sum_{i,j}(J_{i,j}+J_{j,i})h_im_j \cos\theta_j \sin
  \theta_i \cos \theta_i \cos \varphi_i  - \frac{\beta^2}{8}\sum_{i,j}(J_{i,j}+J_{j,i})^2m_j^2
   \cos^2 \theta_j \sin^2 \theta_i  \nonumber \\
  & & + \frac{\beta^2}{4} \sum_{1,j}\left[\left(\frac{1}{4}-m_i^2\right)
  \left(\frac{1}{4}-m_j^2\right) \cos^2 \theta_i \cos^2 \theta_j + \frac{1}{16}
  (1- \cos^2 \theta_i \cos^2\theta_j)
  \right]
\end{eqnarray}
 \end{widetext}
 Eq. (\ref{invrelation}) has been used to eliminate
the dependence from $\lambda_i(0)$. Note that the first line of
the left hand side of Eq. (\ref{GeneralW}) corresponds to the mean
field approximation. In the limit of vanishing transverse field,
i.e. when the symmetry for rotation around the z axis is
recovered, the dependence on $\varphi_i$ disappears as expected.
Moreover if the symmetry breaking fields is fixed along the z -
axis, we recover the free energy expansion of the Ising
model\cite{georges}. This direction is however selected from the
extremum condition on the free energy which corresponds to a
minimum in mean field approximation. In such a case it is easily
shown that, considering a constant interaction among next
neighbors sites, which scales with the system dimensionality
$(J_{i,j}+J_{j,i}=J/D)$, the high temperature expansion becomes an
expansion in $\frac{\beta J}{D}$. This result is actually a
consequence of the compensation of contributions which are of
second order in $\beta$ and zero order in $\frac{1}{D}$. As we
shall see this compensation is valid also in the presence of
transverse field with a suitable choice of the symmetry breaking
fields direction.

Let us consider for simplicity the case of an homogeneous symmetry
breaking $m_i=m$, $\theta_i=\theta$ and $\varphi_i=\varphi$. The
minimum of the free energy in $\varphi$ gives, as expected
$\varphi=0$. The probabilistic functional becomes
 \begin{widetext}
\begin{eqnarray}
 \label{omogeneo}
  W(m,\theta)& = & N\left\{ - \left[ \left(\frac{1}{2}+m\right) \ln\left(\frac{1}{2}+m\right)
  + \left(\frac{1}{2}-m\right) \ln\left(\frac{1}{2}-m\right)
  \right]+ \right.\nonumber \\
 & & \beta \left( \frac{J}{2}m^2 \cos^2\theta + h m \sin \theta \right)
   + \nonumber \\
   & &  \frac{\beta^2}{8} \cos^2 \theta \left[ \left(J m \sin \theta - h
   \right)^2 - \frac{J^2 m^2}{D} \sin^2 \theta \right]+ \nonumber \\
   & & \left. \frac{\beta^2J^2}{4D} \left[ \left(\frac{1}{4}-m^2\right)^2\cos^4 \theta
   + \frac{1}{16} (1 -\cos^4 \theta)
   \right] \right\}
 \end{eqnarray}
 \end{widetext}
Considering only contributions in $\beta$ up to the first order we
recover the well known results of the mean field approximation, by
an extremum condition in $m$ and $\theta$. The first condition
$\frac{\partial W}{\partial m}=0$ gives
\begin{equation}
 \label{derm}
 m = \frac{1}{2} \tanh \left[
 \frac{\beta}{2} \left( J m \cos^2 \theta + h \sin \theta\right) \right]
\end{equation}
corresponds to the vanishing of the symmetry breaking fields, i.e.
to the limit in which the probabilistic generating function $W$
reduces to the free energy. The second corresponds to the
selection of the direction of the symmetry breaking field
corresponding to an extremum of free energy $\frac{\partial
A}{\partial \theta}=0$.
\begin{equation}
 \label{dertheta}
  \cos \theta \left( J m sin\theta -h \right)=0
\end{equation}
It is important to note that the first order expansion in $\beta$
plus the minimum condition in $\theta$, $\varphi$ gives the exact
solution associated to the mean field Hamiltonian. It means that
the rotation in the $\theta$ and $\varphi$ space that diagonalizes
the mean field Hamiltonian coincides with the choice of $\theta$
and $\varphi$ which minimize the first order correction to the
energy.

As far as the second order terms in $\beta$ are concerned, we note
that the ``spurious'' term which does not scale with the inverse
dimensionality $D$, can be neglected close to the magnetization
directions selected in the first order approximation ($\sin
\theta=\frac{h}{J m}$ under the critical temperature, and, $\cos
\theta =0$ above). The solution for the extremum condition, below
the critical temperature for $\theta$ and $m$ modifies as follows:
\begin{equation}
 \label{correctiontheta}
  \sin \theta = \frac{h}{J m}+\frac{\beta h}{4 m J^2 D}
  \left(4h^2 +J^2(1-4m^2) \right)
\end{equation}
\begin{equation}
 \label{correctionm}
 m = \frac{1}{2} \tanh \left\{\frac{\beta}{2} J m - \frac{\beta^2 m}{8 D}
 \left(4h^2 +J^2(1-4m^2) \right) \right\}
\end{equation}
Within this approximation we find an expansion in $\frac{\beta J
}{D}$ which is the generalization of the Ising model
results\cite{georges}. In the limit of vanishing transverse field
we obtain, as expected, the Ising model results.

The critical temperature is identified as the temperature at which
the symmetry breaking direction coincides with that of the
transverse field, i.e $\sin \theta=1$. Solving for the extremum
conditions we obtain a self consistent equation for the critical
temperature.
\begin{equation}
 \label{criticaltemperature}
  \beta_c=\frac{2}{h}arctanh\left(\frac{2h}{J\left(1-\frac{J \beta_c}
  {4D}\right)}\right)
\end{equation}
The solution for Eq. (\ref{criticaltemperature}) is defined in a
limited range of dimensionality and transverse field amplitude.
The lowest critical temperature and highest transverse field are
given by
\begin{eqnarray}
 \label{condizioni critiche}
  \frac{ 2 h}{J \left(1-\frac{\beta J}{4 D}\right)}& = &
   \tanh\left[ D \frac{2 h}{J} \left(1-2 \sqrt{\frac{h^2}{J^2}+\frac{1}{4 D}}
    \right)\right] \nonumber \\
  \beta & = & \frac{4 D}{J} \left(1-2 \sqrt{\frac{h^2}{J^2}+\frac{1}{4 D}} \right)
\end{eqnarray}
In order to compare our results with those derived in reference
(9), we must take into account that this model is the same of Eq.
(\ref{Hamiltonian}) with spin operators substituted by Pauli
operators. This implies the following scaling in the Eqs.
(\ref{omogeneo}) (\ref{criticaltemperature}):$m \rightarrow
\frac{m}{2}$, $J \rightarrow 4J$, $h \rightarrow 2h$.

In Fig. (1) the free energy at critical temperature obtained in
the framework of the standard approach of reference (9) is
compared with our results. We find an improvement for $h\neq0$.
The same comparison for the critical temperature as function of
the transverse magnetic field is performed for dimensionality
greater than four in Fig. (2).
\begin{figure}
\label{energia}
\centerline{\psfig{figure=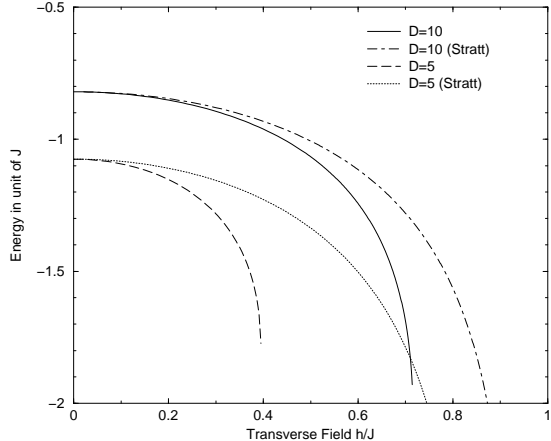,height=6.
cm,angle=-90}} \caption{The free energy for site at critical
temperature as function of transverse field}
\end{figure}
\begin{figure}
\label{temperatura}
 \centerline{\psfig{file=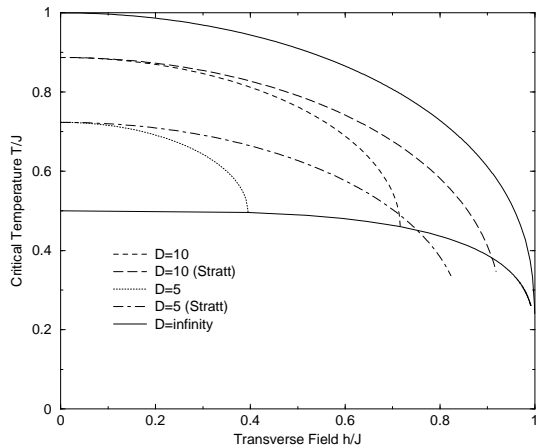,height=6.
cm,angle=-90 }} \caption{The critical temperature as function of
transverse field. The lower curve gives the temperature and the
corresponding transverse field which determines the limit of our
approximation. }
\end{figure}
It is easily seen that the free energy is lowered because of
corrections of order $\frac{1}{D}$ in the direction $\theta$ of
the magnetization. The presence of this extra variational
parameter seems to be the main advantage of the present approach.
As a concluding remark we would like to comment about possible
extensions of this approach to other quantum system. We think
about system defined on a lattice where in the energy there is a
``local'' term, which depends only on the degrees of freedom
associated to a single site, plus an interaction which takes into
account the coupling between degrees of freedom of different
lattice sites. If for a particular choice of the system parameters
there is a symmetry in the local energy operator we can treat
interaction and symmetry breaking part of the local Hamiltonian as
a perturbation. The degeneracy of the system in the limit of
vanishing perturbation, can be removed by the introduction of the
source field, which determines the ''orientation'' in the Hilbert
space of the system ground state. In the case of the quantum Ising
system the local energy is zero and then any local spin state is
allowed. The source field $\lambda$ determines the local state.
The expansion parameter is associated to the inverse temperature
$\beta$. Taking into account the perturbation we can develop an
expansion whose first term is the mean field approximation. The
convergence of this expansion can be improved if a direction
selection mechanism, analogous to that discussed previously
applies. It means that the minimum of the first order
approximation of the free energy (mean field) determines a
direction in the degenerate state Hilbert space and next order
contributions give only small correction to that direction. We are
referring to lattice models where the local energy is determined
by the site particle occupation and particle - particle
interaction on the same site, and the interaction among sites is
associated to hopping of particles from on site to next neighbors
site (Hubbard model). For particular values of the chemical
potential there is a degeneracy of the statistical weight of the
two lowest energy states of the local Hamiltonian. We use the
subspace associated to these states to introduce a general local
state as a superposition of the two independent ground states. The
spontaneous symmetry breakdown will be discussed in terms of the
probability of a local ground state at thermal equilibrium.
Application to repulsive Hubbard model for bosons on a lattice,
and attractive Hubbard model for fermions are in progress.
Preliminary results concerning the boson system can be found in
reference\cite{noi2}.



\begin{thebibliography}{99}

\bibitem{zurek}W. H. Zurek, Phys. Today, 36, {\bf 44}, (1991)
\bibitem{jones}J. A. Jones Prog. NMR Spectrosc, 325, {\bf 38}, (2001).
\bibitem{mattis} D. C. Mattis in ``The Theory of Magnetism II: Thermodynamics
and statistical Mechanics'', Springer Series in Solid - State
Science, {\bf 55} (Springer - Verlag, Berlin, 1985)
\bibitem{wang} Y. L. Wang and B. R. Cooper Phys. Rev. 696 {\bf
185} (1969)
\bibitem{pfeuty} P. Pfeuty and R. J. Elliot, J Phys C, 2370 {\bf
4}, (1971)
\bibitem{weihong}Z. Weihong, J. Oitmaa and C. J. Hamer, J. Phis.
A, 5425, {\bf 27}, (1994)
\bibitem{suzuki} M. Suzuki, Prog.Theor.Phys., 1454, {\bf 56}, (1976)
\bibitem{georges} A. Georges and J.S Yedidia, J.Phys.A:Math.Gen., 2173,
{\bf 24}, (1991)
\bibitem{stratt} R.M. Stratt Phys.Rev.B., 1921, {\bf 33}, (1986)
\bibitem{noi2} F. De Pasquale, S.M. Giampaolo cond-mat/0103293


\end{thebibliography}
\end{document}